\documentclass[conference,onecolumn]{IEEEtran}

\usepackage{cite}
\usepackage{amsmath,amssymb,amsfonts,bm}
\usepackage{algpseudocode}
\usepackage{algorithm}
\usepackage{graphicx}
\usepackage{textcomp}
\usepackage{amsthm}
\usepackage{tikz}
\usepackage{xcolor}
\usepackage{verbatim}
\usepackage{booktabs}
\usepackage{mathtools}
\usepackage{array}
\usepackage{indentfirst}
\usepackage{xpatch}
\usepackage{blindtext}
\usepackage{placeins}
\usepackage{paralist}
\usepackage{nicematrix}
\usepackage{arydshln}
\usepackage[]{footmisc}
\usepackage{standalone}
\usepackage{xfrac}

\usepackage{bbm} 
\usepackage{setspace} 
\usepackage{amsmath}
\usepackage{amssymb}
\usepackage{mathtools}
\usepackage{amsthm}
\usepackage{amsfonts}
\usepackage{nicefrac}
\usepackage{caption}
\usepackage{lipsum}
\usepackage{multirow}
\usepackage{comment}
\usepackage{enumitem}

\ifCLASSOPTIONcompsoc
 \usepackage[caption=false,font=normalsize,labelfont=sf,textfont=sf]{subfig}
\else
 \usepackage[caption=false,font=footnotesize]{subfig}
\fi

\usepackage[switch]{lineno}

\usepackage{dsfont}
\usepackage{diagbox}
\usepackage{booktabs}

\let\epsilon\varepsilon

\let\phi\varphi

\newcommand{\Normal}[2]{\mathcal{N}\!\left(#1, #2\right)}
\newcommand{\Normalx}[3][x]{\mathcal{N}\!\left(#1\, ;\, #2,\, #3\right)}

\renewcommand{\vec}[1]{\mathbf{#1}}
\newcommand{\hatvec}[1]{\hat{\mathbf{#1}}}
\newcommand{\tildevec}[1]{\tilde{\mathbf{#1}}}

\newcommand{\mtrx}[1]{\begin{bmatrix}
    #1
\end{bmatrix}}

\newcommand{\diag}[1]{\text{diag}\left( #1 \right)}
\newcommand{\blkdiag}[1]{\text{blkdiag}\left( #1 \right)}

\newcount\perpcount
\newcommand{\perpp}[1][1]{^{
    \perpcount=#1
    \loop
        \perp
        \advance \perpcount -1
    \ifnum \perpcount>0
    \repeat}
    }
\newcount\adjcount
\newcommand{\adj}[1][1]{^{
    \adjcount=#1
    \loop
        *
        \advance \adjcount -1
    \ifnum \adjcount>0
    \repeat}
    }

\newcommand{\mtext}[2][\qquad]{#1 \text{ #2 } #1}

\newcommand{\inv}[1][1]{^{-#1}}

\newcommand{\R}{\mathbb{R}} 

\newcommand{\sumlim}[2]{\sum_{#1}^{#2}}

\begin{document}
\interdisplaylinepenalty=2500
\title{A Rao-Blackwellized Particle Filter for Superelliptical Extended Target Tracking 
}

 \author{Oğul Can Yurdakul, Mehmet Çetinkaya, Enescan Çelebi, Emre Özkan\\\

 \IEEEauthorblockA{\textit{ Middle East Technical University, Department of Electrical \& Electronics Engineering, Ankara, Turkey } \\
 \texttt{\{ogulyur, cemehmet, enescan.celebi, emreo\}@metu.edu.tr}}}

\maketitle

\bstctlcite{IEEEexample:BSTcontrol}

\begin{abstract}
In this work, we propose a new method to track extended targets of different shapes such as ellipses, rectangles and rhombi. We provide an analytical framework to express these shapes as superelliptical contours and propose a Bayesian filtering scheme that can handle measurements from the contour of the object. The method utilizes the Rao-Blackwellized particle filtering algorithm with novel sensor-object geometry constraints. The success of the algorithm is demonstrated using both simulations and real-data experiments, and the algorithm has been demonstrated to be of high performance in various challenging scenarios. 
\end{abstract}


\section{Introduction}
With the ever-growing number of applications involving closer target-to-sensor proximity and high-resolution sensors, targets can occupy multiple sensor resolution cells, generating multiple measurements.
This motivates the extended target tracking (ETT) problem where both the target extent (shape) and the kinematic states are to be estimated \cite{granstrom2016extended} from these multiple measurements.
ETT algorithms can be used in crowd surveillance, maritime vessel tracking, and tracking both vehicles and pedestrians \cite{granstrom2016extended,granstrom2023tutorial}.
The target extent can be modeled as elliptical \cite{Feldmann_2011 , 8770112 , 10224128}, rectangular \cite{granstrom2014multiple} or star-convex \cite{Wahlström_2015,baum2014extended}.

The random matrix models (RMM) \cite{Koch_2008, Feldmann_2011, Granstrom2014newPred, tuncer2021} are commonly used in ETT algorithms. The underlying assumption is that measurements are distributed across the body of the target. However, this assumption limits their direct use in scenarios with contour measurements \cite{Fowdur2021}, such as those from Lidar sensors. For these cases, algorithms such as the Gaussian process model (GP) \cite{Wahlström_2015, kumru2018} and random hypersurface model (RHS) \cite{baum2009random,baum2014extended} can be used to model complex contours as well as complex surfaces. One approach in \cite{Tuncer2022_MultiEllipse} is to use multiple ellipses to represent the target extent when the measurements are generated from the target contour. 

Another approach for ETT with Lidar measurements is to use geometric object descriptions where the sensor-object geometry can be integrated naturally. In \cite{granstrom2011tracking}, a Gaussian mixture (GM) model probability hypothesis density filter is used to model elliptical and rectangular contours as GMs, where the sensor-object geometry is used to solve the measurement association problem. 
The work by \cite{Michaelis2022} uses a particle filter approach for ETT of a known shape where the expected measurements are obtained using ray casting.
The virtual measurement model (VMM) introduced in \cite{Koch2022_VMM} tracks an elliptical contour using virtual measurements generated via ray tracing.
However, the VMM's virtual measurement generation is computationally expensive, so a later work \cite{KochHoher2023_VMM_GP} trained a Gaussian process regression model to predict the VMM's adaptation law, eliminating the need for virtual measurement generation during runtime.

In this paper, we propose using contours known as superellipses or Lamé curves \cite{Gridgeman1970_LameOvals} to represent the target extent, which are a family of curves containing rhombi, ellipses and more rectangular shapes. 
We choose the Rao-Blackwellized particle filter (RBPF) \cite{doucet2000sequential} as our inference method to explore the potential of using superellipses in extended target tracking. Furthermore, we incorporate the sensor-object geometry into our inference framework analytically, overcoming the challenges of tracking with partially observed contour measurements frequently encountered in Lidar sensor applications. 

\section{Problem Formulation}

\subsection{Superellipses, Extent State and Measurement Model}

\begin{figure*}[ht]
    \centering
    \subfloat[\(q = 1\)]{
        \includegraphics{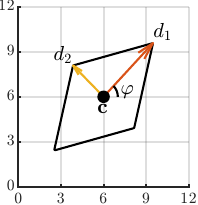}%
        \label{fig:superellipses_q1}
    }
    \hfill
    \subfloat[\(q = 1.5\)]{
        \includegraphics{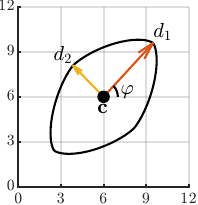}%
        \label{fig:superellipses_q1_5}
    }
    \hfill
    \subfloat[\(q = 2\)]{
        \includegraphics{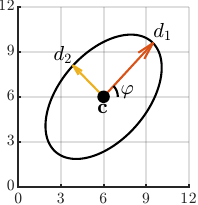}%
        \label{fig:superellipses_q2}
    }
    \hfill
    \subfloat[\(q = 5\)]{
        \includegraphics{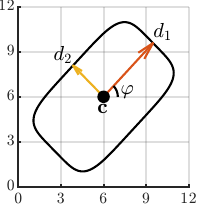}%
        \label{fig:superellipses_q5}
    }
    \caption{Solution sets to \eqref{eqn:superellipse_eqn} for different values of \(q\). In all cases, the extent parameters are selected as \(d^{(1)} = 5\), \(d^{(2)} = 3\), \(\phi = \pi/4\) and \(c^{(1)} = c^{(2)} = 6\). As \(q\) tends to infinity, the contour becomes a rectangle.}
    \label{fig:superellipses}
\end{figure*}

An ellipse in \(\R^2\) can be expressed as the set of points \(\vec y= [y^{(1)} \ y^{(2)}]^T\) that satisfy the equation \begin{equation} \label{eqn:ellipse_Wc}
    (\vec y - \vec c)^T \vec W\inv (\vec y - \vec c) = 1,
\end{equation}
where \(\vec c = [c^{(1)} \ c^{(2)}]^T \in \R^2\) is called the centroid of the ellipse and \(\vec W\) is a symmetric positive-definite matrix. Due to its positive-definiteness, \(\vec W\) can be expressed as \(\vec W = \vec R_\phi \vec D^2 \vec R_\phi^T\), where \(\vec D = \diag{[d^{(1)} \ d^{(2)}]}\) is a diagonal matrix of positive half-lengths along major and minor axes, respectively, and \(\vec R_\phi\) is the rotation matrix by \(\phi\) radians in the counterclockwise direction. With these parameters, \eqref{eqn:ellipse_Wc} can be rewritten as \begin{equation} \label{eqn:ellipse_l2norm_u_y}
    \left(\vec D\inv \vec R_\phi^T (\vec y - \vec c)\right)^T \left(\vec D\inv \vec R_\phi^T (\vec y - \vec c)\right) = 1,
\end{equation}
which implies
\begin{equation} \label{eqn:ellipse_l2norm}
    \|\vec D\inv \vec R_\phi^T (\vec y - \vec c)\|_2^2 = 1,
\end{equation}
where \(\|\cdot\|_2\) denotes the \(\ell_2\) or the Euclidean norm.
One generalization of \eqref{eqn:ellipse_l2norm} can be obtained by not restricting it to the \(\ell_2\) norm, which yields
\begin{equation} \label{eqn:superellipse_eqn}
    \|\vec D\inv \vec R_\phi^T (\vec y - \vec c)\|_q^q = 1,
\end{equation}
where the \(\ell_q\) norm of a vector \(\vec y \in \R^2\) is defined as \begin{equation} \label{eqn:qNorm}
    \left\| \vec y \right\|_q = \left\| \mtrx{y^{(1)} \\ y^{(2)}} \right\|_q = \left(|y^{(1)}|^q + |y^{(2)}|^q\right)^{\nicefrac{1}{q}}.
\end{equation}
for \(q \geq 1\). Note that \eqref{eqn:superellipse_eqn} has a solution for all \(q > 0\), although they do not define a norm for \(1 > q > 0\).
Figure \ref{fig:superellipses} shows the solution sets to \eqref{eqn:superellipse_eqn} for different values of \(q\). These mathematical objects are called Lamé ovals or superellipses when \(q > 2\) \cite{Gridgeman1970_LameOvals}, and have found use in various domains of science, engineering and architecture \cite{gong2004parametric, Kohntopp2019_SuperellipseMine, Duchemin2017_SuperellipseContact, Allen2009_SuperellipseSource}.\footnote{With an abuse of nomenclature, we refer to Lamé curves with all \(q > 0\) as superellipses.}

Consider the variables $\tildevec y = \left[\tilde y^{(1)} \ \tilde y^{(2)}\right]^T$ and \(\boldsymbol \lambda = \left[\lambda^{(1)} \ \lambda^{(2)}\right]^T\), which are defined as
\begin{equation}
    \label{eqn:yTildaCalc}
    \tildevec y := \vec R_\phi^T (\vec y - \vec c)
\end{equation}
and
\begin{equation} \label{eqn:d_to_lambda}
    \lambda^{(j)} = |d^{(j)}|^{-q} \text{ for \(j = 1,2\)},
\end{equation}
respectively. With these definitions, we can write \eqref{eqn:superellipse_eqn} as
\begin{equation} \label{eqn:superellipse_linear}
    \boldsymbol \lambda^T |\tildevec y|^q = \lambda^{(1)} |\tilde y^{(1)}|^q + \lambda^{(2)} |\tilde y^{(2)}|^q = 1,
\end{equation}
where the absolute value and the exponentiation on \(\tildevec y\) are interpreted element-wise. Note that \(\lambda^{(j)}\)'s are strictly positive due to \eqref{eqn:d_to_lambda}.

We can use equation \eqref{eqn:superellipse_linear} to define a non-linear and implicit equation to describe measurements taken from the contour of a superelliptical object. 
When a measurement \(\vec y \in \R^2\) lies exactly on the object contour, we have
\begin{equation} \label{eqn:measEqn_linear}
    0 = h(\vec x^e, \vec y)  = \boldsymbol \lambda^T \left|\vec R_\phi^T (\vec y - \vec c)\right|^q - 1
\end{equation}
satisfied, where
\begin{equation} \label{eqn:extentState_linear}
    \vec x^e := \mtrx{\phi &  c^{(1)} & c^{(2)} & \lambda^{(1)} & \lambda^{(2)} }^T.
\end{equation}
Changing the value of \(q\) changes the contour shape, as illustrated in Figure \ref{fig:superellipses}.

\subsection{Analytical Sensor-Object Geometry Relations}

One problem with extent estimation with contour measurements occurs when the sensor only takes measurements from one side of the target, possibly causing the estimated extent along the unobserved side to deteriorate.
In this section, we derive some analytical conditions for determining which axes of the target are visible. These conditions can be incorporated into the filtering framework to update only the observed extension state estimate.

\begin{figure}[tbp]
    \centering
    \includegraphics{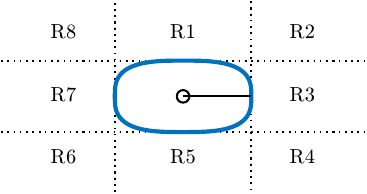}
    \caption{Eight regions in which the sensor could be located with respect to the target. The circle denotes the centroid, the black line denotes the heading axis of the target with \(q = 3\).}
    \label{fig:d_sees}
\end{figure}

\begin{figure*}[tbp]
    \centering
    \subfloat[${[0 \ 3 \ \nicefrac{\pi}{6} \ 2\sqrt{3} \ 2]}^T$]{
        \centering
        \includegraphics{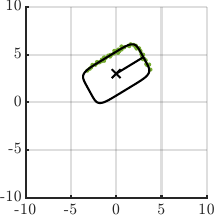}%
        \label{fig:backTransform_y_original}
    }
    \hfill
    \subfloat[${[2 \ (3\text{-}2\sqrt{3}) \ \nicefrac{\pi}{6} \ \sqrt{3} \ 6]}^T$]{
        \centering
        \includegraphics{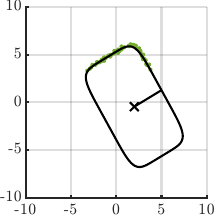}%
        \label{fig:backTransform_y_bad1}
    }
    \hfill
    \subfloat[${[\text{-}3 \ (3\text{-}\sqrt{3}) \ \nicefrac{\pi}{6} \ 2\sqrt{3} \ 2]}^T$]{
        \centering
        \includegraphics{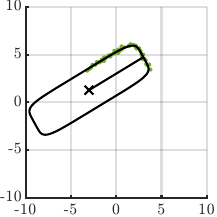}%
        \label{fig:backTransform_y_bad2}
    }
    \hfill
    \subfloat[${[\text{-}2.5 \ 49 \ 0 \ 10 \ 45]}^T$]{
        \centering
        \includegraphics{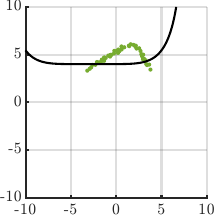}%
        \label{fig:backTransform_y_bad3}
    } 
    \caption{Various superellipses that can fit the point cloud in green, originally generated from the one in \ref{fig:backTransform_y_original} and corrupted with \(\Normal{\vec 0_{2\times 1}}{0.01 \vec I_2}\) noise. The extent parameters \(\vec x^e = {[c^{(1)} \ c^{(2)} \ \phi \ d^{(1)} \ d^{(2)}]}^T\) are given in the captions, with \(q = 6\) is fixed for all superellipses.}
    \label{fig:backTransform}
\end{figure*}

In relation to the target, the sensor can be located in eight different regions as described by tangent lines to the target contour when \(q \geq 2\), located on its four sides. The boundaries of these eight regions are described by the equations
\begin{subequations} \label{eqn:SoG_boundaries}
    \begin{align}
        \left| \vec e_1^T \vec R_\phi^T (\vec y - \vec c) \right| &= d^{(1)} \mtext[\ ]{and} \\
        \left| \vec e_2^T \vec R_\phi^T (\vec y - \vec c) \right| &= d^{(2)},
    \end{align}
\end{subequations}
where $\vec e_j \in \R^2$ is the unit vector along the $j$-th dimension. Each equation in \eqref{eqn:SoG_boundaries} result in two pairs of lines parallel to each other and perpendicular to the other two, as shown in Figure \ref{fig:d_sees} with dotted lines. When the sensor is in regions R2, R4, R6 and R8, both axes of the target are visible, and hence extensions along both axes can be inferred successfully. However, only the extension $d^{(1)}$ along the major axis can be inferred when the sensor is in regions R1 and R5, and that along the minor axis (i.e. $d^{(2)}$) can be inferred when it is in regions R3 and R7. If the sensor location is denoted with \(\vec s = [s^{(1)} \ s^{(2)}]^T\), then its location relative to the target can be computed according to \(\tildevec s := \vec R_\phi^T (\vec s - \vec c)\).
Then, the condition the sensor satisfies in a region where the extension along the dimension \(j\) is visible becomes
\begin{subequations} \label{eqn:SoG_conditions}
    \begin{align}
        \left|\vec e_2^T\vec R_\phi^T (\vec s - \vec c)\right| &> d^{(2)} \qquad \text{for $j = 1$, and} \\
        \left|\vec e_1^T\vec R_\phi^T (\vec s - \vec c)\right| &> d^{(1)} \qquad \text{for $j = 2$}.
    \end{align}
\end{subequations}
These boundaries can be extended by a margin to account for the limited angular sensor resolution as
\begin{subequations} \label{eqn:SoG_conditions_epsilon}
    \begin{align}
        \left|\vec e_2^T\vec R_\phi^T (\vec s - \vec c)\right| &> d^{(2)} + \epsilon^{(2)} \qquad \text{for $j = 1$, and} \\
        \left|\vec e_1^T\vec R_\phi^T (\vec s - \vec c)\right| &> d^{(1)} + \epsilon^{(1)} \qquad \text{for $j = 2$},
    \end{align}
\end{subequations}
which push the boundaries in \eqref{eqn:SoG_boundaries} outwards from the target by \(\epsilon^{(j)} > 0\) along the relevant dimensions. 
For future use, we define the binary variables
\begin{align} \label{eqn:binary_SoG}
    b^{(j)} &:= \mathds{1}\left(\left|\vec e_{j'}^T\vec R_\phi^T (\vec s - \vec c)\right| > d^{(j')} + \epsilon^{(j')}\right) 
\end{align}
to describe the visibility of the object extension along the \(j\)-th axis, where $\mathds{1}(\cdot)$ is the indicator function obtaining the value 1 if the condition in the parentheses is met and 0 otherwise, and \(j' = 2\) when \(j = 1\) and vice versa.

Finally, we note that these conditions are valid for sufficiently rectangular targets, namely those that can be modeled with a high \(q\) value in the superellipse equation \eqref{eqn:superellipse_eqn}. In the case of lower \(q\), especially when \(0 < q \leq 1\), the regions as well as observable properties of the extent contour change.

\subsection{Implicit Measurement with Scale Constraints}

Using the implicit equation \eqref{eqn:measEqn_linear} to find a suitable superellipse fitting a set of measurements \(\mathcal Y = \{\vec y^m\}_{m = 1}^{M}\) requires minimizing the total deviation
    \begin{gather} \label{eqn:superellipse_regression}
        \sumlim{m = 1}{M} \left(h(\vec x^e, \vec y^m)\right)^2 = \sumlim{m = 1}{M} \left(\boldsymbol \lambda^T\left|\vec R_\phi^T (\vec y^m - \vec c)\right|^q - 1\right)^2.
    \end{gather}
Four alternative superellipses that could minimize \eqref{eqn:superellipse_regression} are shown in Figure \ref{fig:backTransform}, where measurements are taken from two sides of the contour in Figure \ref{fig:backTransform_y_original}. Aside from the original superellipse, the other three are also viable options in the sense of \eqref{eqn:superellipse_regression}, as they also have the measurements approximately along their contours in relation to their extent parameters. Notice that if we had measurements from all four sides of the vehicle, the bad fits in Figures \ref{fig:backTransform_y_bad1} and \ref{fig:backTransform_y_bad2} would not occur, while the one in Figure \ref{fig:backTransform_y_bad3} can still occur if the algorithm is initialized poorly.

To prevent the algorithm from converging towards estimates such as in Figures \ref{fig:backTransform_y_bad1} and \ref{fig:backTransform_y_bad2}, we suggest enforcing additional scale constraints based on the transformed measurements $\tildevec y^m := \vec R_\phi^T (\vec y^m - \vec c)$. Assuming no occlusions and sufficiently high sensor resolution, the contour measurements are generated along the full length of each visible side. This means that the transformed measurements \(\tildevec y^m\) will lie exactly in the interval \([-d^{(j)}, d^{(j)}]\) in their \(j\)-th dimension. The limits of the interval are attained regardless of the value of \(q\). That is, for each dimension $j = 1,2$ we have the constraints
\begin{subequations} \label{eqn:scaleConstraints}
    \begin{align} 
        \min_m\left\{\vec e_j^T \vec R_\phi^T (\vec y^m - \vec c)\right\} &= -d^{(j)} \mtext[\ ]{and} \\
        \max_m\left\{\vec e_j^T \vec R_\phi^T (\vec y^m - \vec c)\right\} &= d^{(j)}.
    \end{align}
\end{subequations}
Lastly, it is important to note that each scaling constraint should only be applied if the relevant extension is visible by the sensor according to \eqref{eqn:SoG_conditions_epsilon}.

\section{ETT with Superelliptical Extents} \label{sec:RBPF}

In this section, we introduce our state-space model and our inference method based on the Rao-Blackwellized particle filter (RBPF). In the first subsection, we derive the filtering equations for a superelliptical extent model with a known \(q\) value and extend the formulation to the unknown case in the following subsection.

Let the state-space model with discrete time index \(k\) have the unknown state vector
\begin{align} \label{eqn:stateVector}
    \vec x_{k} := \mtrx{ \phi_{k} & c_{k}^{(1)} & c_{k}^{(2)} & \dot c_k^{(1)} & \dot c_{k}^{(2)} & \lambda_{k}^{(1)} & \lambda_{k}^{(2)} }^T,
\end{align}
where $\dot c^{(1)}_k$ and $\dot c^{(2)}_k$ are the time derivatives of the centroid.
The state dynamics and the measurement equation of the state-space model are defined as
\begin{subequations} \label{eqn:stateSpace}
    \begin{align}
        \vec x_{k+1} & = \vec F\vec x_{k} + \vec G \vec{w}_k, \label{eqn:stateTransition} \\
        0 & = h(\vec x_k, \vec y_k^m) + e_k^m  \label{eqn:measurementEqn}
    \end{align}
    with measurements \(\mathcal Y_k = \{\vec y^m_k\}_{m = 1}^{M_k}\), where
    \begin{equation}
        h(\vec x_k, \vec y_k^m) = \boldsymbol \lambda^T_k\left|\vec R_{\phi_k}^T (\vec y^m_k - \vec c_k)\right|^q - 1. \label{eqn:measurementImplicit}
    \end{equation}
\end{subequations}
$\vec F$ is the state transition matrix, $\vec w_k \sim \mathcal{N}(\vec 0, \vec Q)$ is the process noise and $e_k^m \sim \mathcal{N}(0,r_k)$ is the additive pseudomeasurement noise accounting for the mismatch between the predicted contour and the obtained measurements, describing how much discrepancy from the superellipse equation as given in \eqref{eqn:measEqn_linear} is tolerated.

Consider the filtering problem where we are interested in computing the posterior density $p(\vec x_k|\mathcal{Y}_{1:k})$. Following the RBPF framework, we partition our state variable into the linear state $\vec x_k^l$ and nonlinear state $\vec x_{k}^n$, and decompose their joint posterior as
\begin{subequations}
    \begin{equation}
        p(\vec x_k^l, \vec x_{0:k}^n | \mathcal{Y}_{1:k}) = p(\vec x_k^l | \vec x_{0:k}^n, \mathcal{Y}_{1:k})p(\vec x_{0:k}^n| \mathcal{Y}_{1:k}).
    \end{equation}
The second factor, i.e. the posterior density of nonlinear states, is approximated by an empirical density 
    \begin{equation}
    p(\vec x_{0:k}^n| \mathcal{Y}_{1:k}) \approx \sum_{i=1}^N w_{k,(i)}\delta(\vec x_{0:k}^n - \vec x_{0:k,(i)}^n),
    \end{equation}
with sample trajectories \(\vec x_{0:k,(i)}^n\) and weights \(w_{k,(i)}\).
This approximation allows the conditional density of the linear state given the nonlinear state to be expressed analytically. Under Gaussianity assumptions, the conditional posterior density of the linear state becomes 
    \begin{equation}
        p(\vec x_k^l | \vec x_{0:k,(i)}, \mathcal{Y}_{1:k}) = \Normalx[\vec x_k^l]{\boldsymbol \mu_{k|k,(i)}}{\boldsymbol \Sigma_{k|k,(i)}}.
    \end{equation}
\end{subequations}
where the parameters $\boldsymbol \mu_{k|k,(i)}$ and $\boldsymbol \Sigma_{k|k,(i)}$ are the conditional posterior mean and covariance, respectively, which can be computed using a Kalman filter \cite{kalman1960new} run conditionally on the nonlinear state.

\subsection{Superelliptical Extent Model with Known \(q\)}

Let $\vec x_k^n:=[\phi_{k} \ c_{k}^{(1)} \ c_{k}^{(2)} \ \dot c_k^{(1)} \ \dot c_{k}^{(2)}]^T$ denote our nonlinear state, while \(q\) is assumed to be known. Conditioned on $\vec x_k^n$, the measurement equation in \eqref{eqn:measurementEqn} is linear in the remaining states $\vec x_k^l:=\boldsymbol \lambda_k$. Therefore, we partition \cite{Schon2005} the state dynamics equation \eqref{eqn:stateTransition} as
\begin{subequations} \label{eqn:stateMatrices}
\begin{align}
    \vec x_{k+1}^{n} & = \vec F^n \vec x_k^n +  \vec G^n \vec w_k^n, \label{eqn:stateTransitionNL} \\
    \vec x_{k+1}^{l} & = \vec x_k^l + \vec w_k^l, \label{eqn:stateTransitionL}
\end{align}
\end{subequations} 
with
\begin{subequations} \label{eqn:stateDynamics}
    \begin{align}
        \vec F^n &= \text{blkdiag}\left(1, \mtrx{1 & T \\ 0 & 1} \otimes \vec I_2\right) \\
        \vec G^n &= \text{blkdiag}\left(1, \mtrx{\nicefrac{T^2}{2} \\ T} \otimes \vec I_2\right)
    \end{align}
    \begin{alignat}{2}
        \vec w_k^n &\sim \Normal{\vec 0_{3 \times 1}}{\vec Q^n} \text{ with } && \vec Q^n = \text{blkdiag}\left(\sigma_\phi^2, \sigma_a^2 \vec I_2\right) \\
        \vec w_k^l &\sim \Normal{\vec 0_{2 \times 1}}{\vec Q^l} \text{ with } && \vec Q^l = \sigma^2_\lambda \vec I_2
    \end{alignat}
\end{subequations} 
where $\vec w_k^n$ is the vector of angle and acceleration noises of the constant velocity model, $\vec w_k^l$ is the extension process noise, $\vec I_n$ denotes the $n \times n$ identity matrix and $\otimes$ denotes the Kronecker product.

In the RBPF framework, the linear state measurement update (KF-MU) and time update (KF-TU) steps are performed analytically conditioned on particles representing the nonlinear states. For the nonlinear states, the weight update (PF-WU) and sampling (PF-S) steps are performed. We introduce the details of these steps in the following. A pseudocode for this filter is given in Algorithm \ref{alg:RBPF}.

\paragraph{PF-WU}
The weight update of the particles can be expressed as a multiplication of the previous weights with corresponding incremental weights, denoted as $\alpha_{k,(i)}$. Using \eqref{eqn:measurementEqn}, the incremental pseudomeasurement update weights are 
\begin{gather} \label{eqn:PF_likelihood}
    \alpha^h_{k,(i)} =  \prod_{m = 1}^{M_k} \Normalx[\boldsymbol \mu^T_{k|k-1,(i)} \left|\vec R_{\phi_{k,(i)}}^T (\vec y^m_k - \vec c_{k,(i)})\right|^q]{1}{r_k^h},
\end{gather} 
where \(r_k^h\) is the covariance.
\begin{subequations}
    The marginalized covariance becomes
    \begin{equation}
        r_k +  \left(\left|\vec R_{\phi_{k,(i)}}^T (\vec y^m_k - \vec c_{k,(i)})\right|^q\right)^T \vec \Sigma_{k|k-1,(i)} \left|\vec R_{\phi_{k,(i)}}^T (\vec y^m_k - \vec c_{k,(i)})\right|^q.
    \end{equation}
    To have additional control over the discrepancy from the superellipse equation, we set \(r_k^h\) for each particle as 
    \begin{equation}
        r_k^h = \gamma_{k,(i)}^m\left(r_k +  \left(\left|\vec R_{\phi_{k,(i)}}^T (\vec y^m_k - \vec c_{k,(i)})\right|^q\right)^T \vec \Sigma_{k|k-1,(i)} \left|\vec R_{\phi_{k,(i)}}^T (\vec y^m_k - \vec c_{k,(i)})\right|^q\right),
    \end{equation}
    where \(\gamma_{k,(i)}^m\) is a positive scaling term.
\end{subequations}

If the extension along the $j$-th axis is deemed to be visible according to \eqref{eqn:SoG_conditions_epsilon}, its corresponding scale constraints \eqref{eqn:scaleConstraints} can be applied as a soft constraint using Gaussian factors. For each dimension $j = 1,2$ the corresponding scale constraint in \eqref{eqn:scaleConstraints} is expressed as
\begin{subequations} \label{eqn:constraint_alpha}
    \begin{align}
        \alpha^{s,(j)}_{k,(i)} &= \Normalx[\min_m\left\{\vec e_j^T \vec R_{\phi_{k,(i)}}^T (\vec y^m_k - \vec c_{k,(i)})\right\}]{-\hat d^{(j)}_{k,(i)}}{r_k^s} \\
        & \quad \times \Normalx[\max_m\left\{\vec e_j^T \vec R_{\phi_{k,(i)}}^T (\vec y^m_k - \vec c_{k,(i)})\right\}]{\hat d^{(j)}_{k,(i)}}{r_k^s},
    \end{align}
    where $r_k^s$ is the variance parameter for the scaling constraint, and $\hat d^{(j)}_{k,(i)}$ is calculated using $\mu_{k|k-1, (i)}^{(j)} = \hat \lambda_{k|k-1, (i)}^{(j)}$ from the relation between $d^{(j)}$ and $\lambda^{(j)}$ according to \eqref{eqn:d_to_lambda}. The corresponding binary variables \(b^{(j)}_{k,(i)}\) for the sensor-object geometry criteria along each dimension are calculated according to \eqref{eqn:binary_SoG}.
    The combined scale constraint for both axes including the sensor-object geometry criteria can be expressed as
    \begin{equation} \label{eqn:incWeightUpdate}
        \alpha^{s}_{k,(i)} = \left(\alpha^{s,(1)}_{k,(i)}\right)^{b^{(1)}_{k,(i)}}\left(\alpha^{s,(2)}_{k,(i)}\right)^{b^{(2)}_{k,(i)}},
    \end{equation}
\end{subequations} 
which results in the weight update
\begin{equation} \label{eqn:unnormalized_weights}
    \tilde w_{k,(i)} = w_{k-1,(i)} \times \alpha^h_{k,(i)} \times \alpha^{s}_{k,(i)} \times 
        \frac{
            p\left( \vec x^n_{k,(i)} | \vec x^n_{0:k-1,(i)}, \mathcal Y_{1:k-1} \right)
        }{
            q\left( \vec x^n_{k,(i)} | \vec x^n_{0:k-1,(i)}, \mathcal Y_{1:k} \right)
        }.
\end{equation}
where $q\left( \vec x^n_{k,(i)} | \vec x^n_{0:k-1,(i)}, \mathcal Y_{1:k} \right)$ is the proposal density from which the particles are sampled.
Finally, to ensure the positivity of the extent variables, the weights of the particles with negative $\lambda_{k,(i)}^{(j)}$ are set to zero.

\paragraph{KF-MU}
Conditional on the nonlinear states, the measurement equation in \eqref{eqn:measurementEqn} can be compactly written as
\begin{subequations} \label{eqn:KF_measEqn}
    \begin{equation}
        \label{eqn:pseudoMeasurementEqn}
        \vec 1_{M_k \times 1} = \vec H_k(\vec x_{k,(i)}^n,\mathcal{Y}_k) \vec x_k^l + \vec e_k,
    \end{equation}
where $\vec 1_{M_k \times 1}$ denotes the $M_k \times 1$ matrix of ones, $\vec e_k\sim\mathcal{N}(\vec 0_{M_k \times 1},r_k \vec I_{M_k})$ and
    \begin{align} \label{eqn:KF_MU_H}
        \vec H_k(\vec x_{k,(i)}^n,\mathcal{Y}_k) = \mtrx{ \left(\left|\vec R_{\phi_{k,(i)}}^T (\vec y^1_k - \vec c_{k,(i)})\right|^q\right)^T \\ \vdots \\ \left(\left|\vec R_{\phi_{k,(i)}}^T (\vec y^{M_k}_k - \vec c_{k,(i)})\right|^q\right)^T },
    \end{align}
\end{subequations}
where $m = 1,...,M_k$. Then, omitting the dependence of $\vec H_k(x_{k,(i)}^n,\mathcal{Y}_k)$ on $\vec x_{k,(i)}^n$ and $\mathcal{Y}_k$ for notational brevity, the Kalman filter measurement update equations can be written as
\begin{subequations} \label{eqn:KF_MU}
    \begin{align}
        \boldsymbol \mu_{k|k,(i)}    & = \boldsymbol\mu_{k|k-1,(i)} + \vec K_{k,(i)}(\vec 1_{M_k \times 1}-\vec H_k \boldsymbol \mu_{k|k-1,(i)}),                    \label{eqn:Kalman_mean}\\
        \boldsymbol \Sigma_{k|k,(i)} & = \boldsymbol \Sigma_{k|k-1,(i)} - \vec K_{k,(i)} \vec S_{k,(i)} \vec K_{k,(i)}^T, \label{eqn:Kalman_cov}
        \intertext{where}
        \vec S_{k,(i)}               & = \vec H_k \boldsymbol \Sigma_{k|k-1,(i)} \vec H_k^T + r_k \vec I_{M_k},                                                    \\
        \vec D_{k,(i)}                 & = \diag{b^{(1)}_{k,(i)}, b^{(2)}_{k,(i)}}, \label{eqn:K_mask}\\
        \vec K_{k,(i)}               & = \vec D_{k,(i)} \boldsymbol \Sigma_{k|k-1,(i)} \vec H_k^T \vec S_{k,(i)}\inv. \label{eqn:masked_K}
    \end{align}
\end{subequations}
Notice that \eqref{eqn:masked_K} would be the standard Kalman gain, except for the \(\vec D_{k, (i)}\) term with two diagonal entries given in \eqref{eqn:binary_SoG}. 
This matrix, which we refer to as the Kalman gain mask, multiplies the rows of the original Kalman gain matrix by zero if the corresponding extent is not visible. 
This allows the mean and variance of that extent variable to remain unchanged after the update equations in \eqref{eqn:Kalman_mean} and \eqref{eqn:Kalman_cov}. 
While such an update results in a drift from a fully Bayesian approach, it enables analytical incorporation of the sensor-object geometry constraints into our filtering framework easily.

\paragraph{PF-S}
To perform the particle filter time update step, we draw new particles from the proposal density
\begin{gather} \label{eqn:PF_TU_sampling}
     \vec x_{k+1,(i)}^n \sim q\left( \vec x^n_{k+1,(i)} | \vec x^n_{0:k,(i)}, \mathcal Y_{1:k} \right).
\end{gather}
If the proposal density is selected as the transition density $p\left( \vec x^n_{k+1,(i)} | \vec x^n_{0:k,(i)}, \mathcal Y_{1:k} \right)$, the samples will be drawn from the following Gaussian proposal
\begin{align}
    q(\cdot) = \Normalx[\vec x_{k+1}^n]{\vec F^n \vec x_{k,(i)}^n}{\vec G^n\vec Q^n (\vec G^n)^T}. \label{eqn:PF_proposal}
\end{align}

\paragraph{KF-TU}
Finally, the Kalman filter time update is performed in accordance with the dynamics in \eqref{eqn:stateTransitionL}, as
\begin{subequations} \label{eqn:KF_TU_pmeas_TU}
    \begin{align}
        \boldsymbol \mu_{k+1|k,(i)}    & = \boldsymbol \mu_{k|k,(i)}, \\
        \boldsymbol \Sigma_{k+1|k,(i)} & = \boldsymbol \Sigma_{k|k,(i)} + \vec Q^l.
    \end{align}
\end{subequations}

\begin{algorithm}[tbp]
    \setstretch{1.25}
    \begin{algorithmic}[1] 
        \State Initialize particles by sampling $\vec x_{1,(i)}^n$ from the prior with identical weights $w_{1,(i)} = 1/N$, conditional  means \(\boldsymbol \mu_{1|0,(i)}\) and covariances \(\boldsymbol \Sigma_{1|0,(i)}\) for all \(i = 1,...,N\).
        \For {$k = 1:K$}
        \For {$i = 1:N$} \Comment PF-WU\footnotemark
            \State Compute \(\tilde w_{k,(i)}\) using \eqref{eqn:PF_likelihood} - \eqref{eqn:unnormalized_weights} 
        \EndFor
        \State Normalize weights \(w_{k,(i)} = \tilde w_{k,(i)} / \sumlim{i' = 1}{N} \tilde w_{k,(i')}\)
        \State Resample particles with replacement 
        \For {$i = 1:N$} \Comment KF-MU, PF-S \& KF-TU
            \State Compute \(\boldsymbol \mu_{k|k,(i)}\) and \(\boldsymbol \Sigma_{k|k,(i)}\) using \eqref{eqn:KF_measEqn} - \eqref{eqn:KF_MU} 
            \State Propagate \(\vec x_{k,(i)}^n\) to \(\vec x_{k+1,(i)}^n\) using \eqref{eqn:PF_TU_sampling} 
            \State Compute \(\boldsymbol \mu_{k+1|k,(i)}\) and \(\boldsymbol \Sigma_{k+1|k,(i)}\) using \eqref{eqn:KF_TU_pmeas_TU} 
        \EndFor
        \EndFor
    \end{algorithmic}
    \caption{A Rao-Blackwellized Particle Filter for ETT with Superelliptical Extents}
    \label{alg:RBPF}
\end{algorithm} 

\subsection{Superelliptical Extent Model with Unknown \(q\)} \label{sec:varQ}

The algorithm can be easily modified to estimate the unknown shape of an object by defining the superellipse exponent $q$ as an unknown variable. This allows the algorithm to adapt to the target shape with a suitable superellipse. In this case, the nonlinear state vector can be augmented with \(q_k\), i.e. $\vec x_k^n:=[q_k \ \phi_{k} \ c_{k}^{(1)} \ c_{k}^{(2)} \ \dot c_k^{(1)} \ \dot c_{k}^{(2)}]^T$, while the linear state $\vec x_k^l:=\boldsymbol \lambda_k$ is kept the same. 
Then, the nonlinear state dynamics parameters in \eqref{eqn:stateDynamics} are modified to be 
\begin{subequations}
    \begin{align}
        \vec F^n &= \text{blkdiag}\left(\vec I_2, \mtrx{1 & T \\ 0 & 1} \otimes \vec I_2\right), \\
        \vec G^n &= \text{blkdiag}\left(\vec I_2, \mtrx{\nicefrac{T^2}{2} \\ T} \otimes \vec I_2\right), 
    \end{align}
    \begin{align}
        \vec w_k^n &\sim \Normal{\vec 0_{4 \times 1}}{\vec Q^n} \text{ with } \vec Q^n = \text{blkdiag}\left(\sigma_q^2, \sigma_\phi^2, \sigma_a^2 \vec I_2\right),
    \end{align}
\end{subequations} 
with the \(q_k\) variable assumed to be slowly varying together with the strict positivity constraint.

\footnotetext{For numerical stability, log-likelihoods are used and the weights are multiplied by appropriate constants when necessary.}

\section{Experimental Results}
The performance of the algorithm is evaluated in four different simulation scenarios and one scenario with real data. 
In the scenarios where \(q\) is known, the nonlinear states $\vec x_{1,(i)}^n$ are sampled from the Gaussian \(\Normal{\boldsymbol \mu^n_{0}}{\boldsymbol \Sigma^n_{0}}\) with 
\begin{subequations}
    \begin{align}
        \boldsymbol \mu^n_{0} &= \mtrx{0 & \bar m^{(1)} & \bar m^{(2)} & \bar v^{(1)} & \bar v^{(2)}}^T, \\
        \boldsymbol \Sigma^n_{0} &= \blkdiag{\left(\nicefrac{\pi}{4}\right)^2, \vec I_2, 4 \vec I_2},
        \intertext{where $\bar m^{(j)}$ are the center coordinates of the bounding box containing the first set of measurements and \(\bar v^{(j)}\) is the velocity estimate obtained via the means of the first two sets of measurements. We follow this two-step initialization approach for the velocity component to improve the particle efficiency. In the known shape scenarios, the superellipse exponent \(q\) is selected as \(q = 5\). Each particle's conditional mean and variance are set to}
        \boldsymbol \mu_{1|0,(i)} &= \mtrx {2^{-q} & 1^{-q}}^T \mtext[\ ]{and} \boldsymbol \Sigma_{1|0,(i)} = 100 \vec I_2
    \end{align}
\end{subequations}
for \(i = 1, ..., N\) where $N = 1000$ is the number of particles. Notice that this selection of \(\boldsymbol \mu_{1|0,(i)}\) initializes the extent half-lengths as \(2\ m\) and \(1\ m\).

Throughout all simulations, the process noise standard deviations are selected as $\sigma_\phi = \nicefrac{\pi}{36}$, $\sigma_a = 2$ and $\sigma_\lambda = 10^{-4}$, 
while the proposal density is chosen as in \eqref{eqn:PF_proposal}. 
The extension observation conditions in \eqref{eqn:SoG_conditions_epsilon} are used with $\epsilon^{(1)} = \epsilon^{(2)} = 0.5$ $m$.
The implicit measurement covariance in \eqref{eqn:PF_likelihood} and scale constraint covariance in \eqref{eqn:constraint_alpha} are selected as $r_k=r_k^s=(0.3)^2$, while \(\gamma_{k,(i)}^m\) is selected to set \(r_k^h=(0.3)^2\) for all particles and measurements.

\subsection{Simulation Results}

\begin{figure}[tbp]
    \centering
    \subfloat[Curved scenario, plotted every 25 frames.]{
        \centering
        \includegraphics[width=0.35\linewidth]{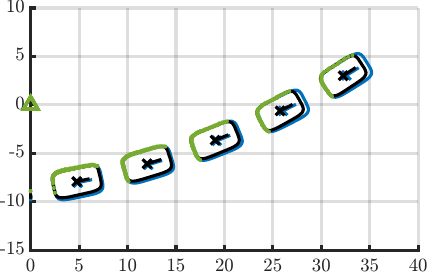}%
        \label{fig:Curved_fixQ}
    }
    \hfill
    \subfloat[Drifting scenario, plotted every 25 frames.]{
        \centering
        \includegraphics[width=0.35\linewidth]{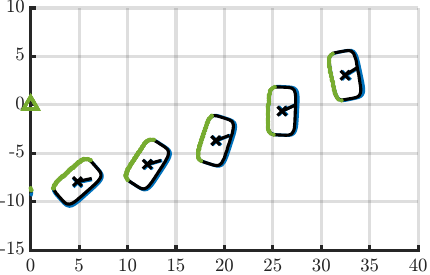}%
        \label{fig:Drifting_fixQ}
    }
    \hfill
    \subfloat[U-turn scenario, plotted every 35 frames.]{
        \centering
        \includegraphics[width=0.25\linewidth]{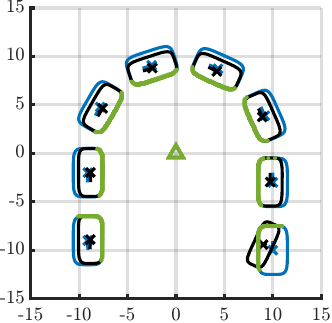}%
        \label{fig:UTurn_fixQ}
    }
    \caption{Sample outputs of the algorithm with known \(q = 5\). The solid black lines indicate the posterior (contour, centroid and velocity vector) while the blue ones indicate the true extent (contour, centroid and velocity vector). The triangle indicates the sensor location.}
    \label{fig:outputs}
\end{figure}

We test the performance of our algorithm in the following four simulated scenarios. 
\begin{description}
    \item[\textbf{Linear,}] where the target follows a linear trajectory along its main axis without drifting, starting from (-40, -10) and ending at (35, -10) with a linear velocity of $3$ $\nicefrac{m}{s}$,
    \item[\textbf{Curved,}] where the target follows a similar trajectory starting from (-40, -10), except for a slight turn towards the positive direction (Figure \ref{fig:Curved_fixQ}),
    \item[\textbf{Drifting,}] where the trajectory is similar to that of the Curved scenario, except the target also starts drifting while turning (Figure \ref{fig:Drifting_fixQ}), and
    \item[\textbf{U-turn,}] where the target follows a semicircle with a radius of $10$ $m$ centered around the sensor between brief periods of linear motion, with a linear velocity of $2$ $\nicefrac{m}{s}$ (Figure \ref{fig:UTurn_fixQ}).
\end{description}

Roughly ordered in an increasingly challenging order, each of these scenarios serves to test the performance of the algorithm in different respects. The first two are relatively simple scenarios to demonstrate elementary performance. The latter two illustrate performance in cases where the second extension of the target is not visible for a significant amount of time, testing the effectiveness of gain masking in \eqref{eqn:K_mask}.

Throughout the simulations, the sampling time is chosen as $T=0.1 \ s$. The underlying true shape is a superellipse with \(q = 5\) and half-lengths $d^{(1)}=2.5 \ m$ and $d^{(2)}=1.5 \ m$.
The measurement noise is simulated by adding zero-mean Gaussian noises to the range and bearing of the noise-free Lidar measurements, with the standard deviations set to $\sigma_r = 0.01m$ and $\sigma_\theta = 0.005^\circ$, respectively.
The Lidar resolution is chosen to be $0.2^\circ$ with a \(360^\circ\) field of view and it is assumed to be positioned at the origin.

\begin{table}[tbp]
    \centering
    \caption{Performance metrics and computation times in different simulation scenarios with known shape. All results are averages over 100 Monte Carlo runs.}
    \begin{tabular}{@{}cccccc@{}}
    \toprule
    \multicolumn{2}{c}{\textbf{Scenario}}                         & {Linear} & {Curved} & {Drifting} & {U-turn} \\ \midrule
    \multirow{4}{*}{\rotatebox{90}{\textbf{RMSE}}}& $\vec c\ (m)$ & 0.3302 & 0.3694 & 0.3249 & 0.3728 \\ \cmidrule{2-6}
    & $\vec v\ (\nicefrac{m}{s})$                                 & 0.2744 & 0.3217 & 0.2978 & 0.3145 \\ \cmidrule{2-6}
    & $d^{(1)}\ (m)$                                              & 0.2139 & 0.2247 & 0.2032 & 0.1383 \\ \cmidrule{2-6}
    & $d^{(2)}\ (m)$                                              & 0.1340 & 0.2193 & 0.1757 & 0.3055 \\ \midrule
    \multicolumn{2}{c}{\textbf{IOU}}                              & 0.8297 & 0.8202 & 0.6846 & 0.7609 \\ \midrule
    \multicolumn{2}{c}{\textbf{Time} $(s)$}                       & 0.2023 & 0.2280 & 0.2314 & 0.3363 \\ \bottomrule
    \end{tabular}
    \label{tab:rmse_table}
\end{table}

We use root mean square error (RMSE) and Intersection-Over-Union (IOU) \cite{alexe2012measuring} as our performance metrics over $M_c=100$ Monte Carlo runs.
RMSEs of the state estimates are
\begin{equation}
    \text{RMSE}_{x} = \sqrt{\frac{1}{KM_c} \sum_{k=1}^K \sum_{i=1}^{M_c} (x_k-\hat{x}_{k}^i)^2 },
\end{equation}
where $x_k$ denotes the true state and $\hat{x}_{k}^i$ is its estimate, calculated according to
\begin{align} \label{eqn:stateEstimates}
    \hatvec x_{k}^n = \sumlim{i = 1}{N} w_{k,(i)}\vec x_{k,(i)}^n, \mtext[\ ]{and}
    \hatvec x_{k}^l = \sumlim{i = 1}{N} w_{k,(i)}\boldsymbol \mu_{k|k,(i)}.
\end{align} Velocity and centroid errors are calculated using their respective Euclidean norms. To account for the performance in extent estimation, we calculate IOU according to
\begin{equation}
    \text{IOU} = \frac{1}{KM_c} \sum_{k=1}^K \sum_{i=1}^{M_c} \frac{\text{area}\left( \hatvec X^e_{k,i} \cap \vec X^e_{k,\text{true}} \right)}{\text{area}\left( \hatvec X^e_{k,i} \cup \vec X^e_{k,\text{true}} \right)}
\end{equation}
where $\hatvec X^e_{k,i}$ denotes the posterior superellipse region described by the posterior extent state $\hatvec x^e_{k,i}$ in Monte Carlo run $i$, and $\hatvec X^e_{k,\text{true}}$ is the true extent of the target. 

\subsubsection{Known Shape Scenarios}

Sample outputs of the algorithm with a known shape, i.e. a known \(q\) value, are given in Figure \ref{fig:outputs} while the performance metrics and computation times are given in Table \ref{tab:rmse_table}. 
Note that the RMSEs are comparable in the U-turn and Drifting scenarios, even when one side of the target is invisible throughout most of the simulation. The half-length state corresponding to \(d^{(2)}\), unobservable during these scenarios, also exhibits sufficiently small RMSE values in these scenarios, indicating that the corresponding state estimate has not deteriorated.
These show the effectiveness of the Kalman gain mask $\vec D_{k,(i)}$, preventing erroneous updates to the extension estimates and preserving the centroid estimate.

\subsubsection{Unknown Shape Scenario}

\begin{figure}[tbp]
    \centering
    \includegraphics{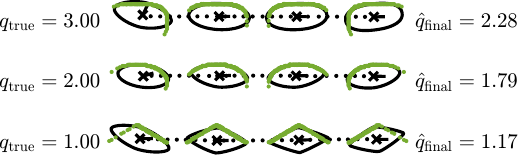}%
    \caption{Three different simulation results with the augmented target state. The vehicle is moving from left to right. Solid lines indicate the estimated contour and the velocity vector, the dotted line indicates the estimated centroid trajectory and the final and true \(q\) values are given in the figure.}
    \label{fig:variableQ}
\end{figure}

To illustrate the performance of the algorithm for targets with unknown shapes, i.e. unknown \(q\), we run the algorithm with the augmented state in Section \ref{sec:varQ} in three elementary scenarios. 
In these scenarios, targets follow a linear trajectory with constant speed over $K = 100$ time steps. The state \(q_k\) is initialized by sampling from the prior \(q_{1,(i)}\sim\Normal{2}{(0.2)^2}\) with its process noise standard deviation set to \(\sigma_q = 0.1\), while \(N = 1500\) particles are used to compensate for the increase in the nonlinear state dimension. All other parameters are left identical. The results are presented in Figure \ref{fig:variableQ}. It is seen that the algorithm can find a superellipse that accommodates the obtained measurements. The estimated centroids also appear to reflect a suitable trajectory for the target. 

\subsection{Real Data Application} 

\begin{figure}[tbp]
    \centering
    \subfloat[Known shape with \(q = 5\).]{
        \centering
        \includegraphics{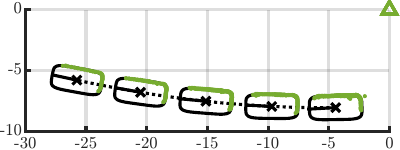}%
        \label{fig:RealData_fixQ}
    }
    \qquad
    \subfloat[Unknown shape initialized with an ellipse.]{
        \centering
        \includegraphics{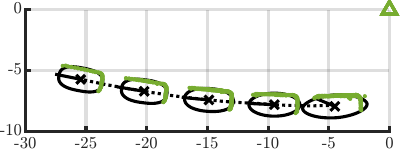}%
        \label{fig:RealData_varQ}
    }
    \caption{Example output of the algorithm with real data from \cite{Tuncer2022_MultiEllipse}, plotted every 12 frames. The vehicle is moving from right to left. The solid black lines indicate the posterior (contour, centroid and velocity vector) while the dotted line indicates the centroid trajectory.}
    \label{fig:realData_output}
\end{figure}

Finally, we illustrate the performance of our algorithm on real data. For this purpose, we use the data presented in \cite{Tuncer2022_MultiEllipse}, where a Lidar sensor with $T = 0.1\ s$ sampling time and angular resolution $0.2^\circ$ is used to collect contour measurements from a moving vehicle whose rear and right sides are visible. 
On the same dataset, we present the output of both the known and unknown shape approaches.
For the known shape algorithm, the superelliptical exponent \(q = 5\) is assumed to better represent the target as an automobile,
while the unknown shape algorithm is initialized again with \(\Normal{2}{(0.2)^2}\) for the prior of the state \(q_k\).
The outputs of the algorithms are shown in Figure \ref{fig:realData_output}. Both algorithms successfully track the vehicle and estimate its velocity, and the algorithm with the adaptable target shape finds a suitable superellipse although initialized with an elliptical prior.

\section{Conclusion}
In this paper, we have proposed a new ETT algorithm for contour measurements which represents the target shape as a family of curves known as superellipses. Through the variation of a single exponent parameter, the algorithm unifies a variety of shapes such as diamonds, ellipses and rectangles with varying degrees of rounded corners in a single, implicit measurement equation. We have used this model in a Rao-Blackwellized particle filter, with scaling constraints to prevent particles from reaching local minima of the implicit measurement model proposed and analytical sensor-object geometry relations by modifying the Kalman filter equations. Using both simulations and real-world data, we have shown that the algorithm has high tracking performance, as well as being able to estimate the unknown shape of the target.

\appendices 

\bibliographystyle{IEEEtran}
\bibliography{IEEEabrv,references}

\begin{thebibliography}{10}
\providecommand{\url}[1]{#1}
\csname url@samestyle\endcsname
\providecommand{\newblock}{\relax}
\providecommand{\bibinfo}[2]{#2}
\providecommand{\BIBentrySTDinterwordspacing}{\spaceskip=0pt\relax}
\providecommand{\BIBentryALTinterwordstretchfactor}{4}
\providecommand{\BIBentryALTinterwordspacing}{\spaceskip=\fontdimen2\font plus
\BIBentryALTinterwordstretchfactor\fontdimen3\font minus
  \fontdimen4\font\relax}
\providecommand{\BIBforeignlanguage}[2]{{%
\expandafter\ifx\csname l@#1\endcsname\relax
\typeout{** WARNING: IEEEtran.bst: No hyphenation pattern has been}%
\typeout{** loaded for the language `#1'. Using the pattern for}%
\typeout{** the default language instead.}%
\else
\language=\csname l@#1\endcsname
\fi
#2}}
\providecommand{\BIBdecl}{\relax}
\BIBdecl

\bibitem{granstrom2016extended}
K.~Granstrom, M.~Baum, and S.~Reuter, ``Extended object tracking: Introduction,
  overview and applications,'' \emph{arXiv preprint arXiv:1604.00970}, 2016.

\bibitem{granstrom2023tutorial}
K.~Granstr{\"o}m and M.~Baum, ``A tutorial on multiple extended object
  tracking,'' \emph{Authorea Preprints}, 2023.

\bibitem{Feldmann_2011}
M.~Feldmann, D.~Fränken, and W.~Koch, ``Tracking of extended objects and group
  targets using random matrices,'' \emph{IEEE Transactions on Signal
  Processing}, 2011.

\bibitem{8770112}
S.~Yang and M.~Baum, ``Tracking the orientation and axes lengths of an
  elliptical extended object,'' \emph{IEEE Transactions on Signal Processing},
  vol.~67, no.~18, pp. 4720--4729, 2019.

\bibitem{10224128}
S.~Steuernagel, K.~Thormann, and M.~Baum, ``Improved extended object tracking
  with efficient particle-based orientation estimation,'' in \emph{2023 26th
  International Conference on Information Fusion (FUSION)}.\hskip 1em plus
  0.5em minus 0.4em\relax IEEE, 2023, pp. 1--8.

\bibitem{granstrom2014multiple}
K.~Granstr{\"o}m, S.~Reuter, D.~Meissner, and A.~Scheel, ``A multiple model
  {PHD} approach to tracking of cars under an assumed rectangular shape,'' in
  \emph{17th International Conference on Information Fusion (FUSION)}.\hskip
  1em plus 0.5em minus 0.4em\relax IEEE, 2014, pp. 1--8.

\bibitem{Wahlström_2015}
N.~Wahlstr{\"o}m and E.~{\"O}zkan, ``Extended target tracking using {G}aussian
  processes,'' \emph{IEEE Transactions on Signal Processing}, vol.~63, no.~16,
  pp. 4165--4178, 2015.

\bibitem{baum2014extended}
M.~Baum and U.~D. Hanebeck, ``Extended object tracking with random hypersurface
  models,'' \emph{IEEE Transactions on Aerospace and Electronic systems},
  vol.~50, no.~1, pp. 149--159, 2014.

\bibitem{Koch_2008}
J.~W. Koch, ``{B}ayesian approach to extended object and cluster tracking using
  random matrices,'' \emph{IEEE Transactions on Aerospace and Electronic
  Systems}, vol.~44, no.~3, pp. 1042--1059, 2008.

\bibitem{Granstrom2014newPred}
K.~Granstr{\"o}m and U.~Orguner, ``New prediction for extended targets with
  random matrices,'' \emph{IEEE Transactions on Aerospace and Electronic
  Systems}, vol.~50, no.~2, pp. 1577--1589, 2014.

\bibitem{tuncer2021}
B.~Tuncer and E.~{\"O}zkan, ``Random matrix based extended target tracking with
  orientation: A new model and inference,'' \emph{IEEE Transactions on Signal
  Processing}, vol.~69, pp. 1910--1923, 2021.

\bibitem{Fowdur2021}
J.~S. Fowdur, M.~Baum, and F.~Heymann, ``An elliptical principal axes-based
  model for extended target tracking with marine radar data,'' in \emph{2021
  IEEE 24th International Conference on Information Fusion (FUSION)}.\hskip 1em
  plus 0.5em minus 0.4em\relax IEEE, 2021, pp. 1--8.

\bibitem{kumru2018}
M.~Kumru and E.~{\"O}zkan, ``3{D} extended object tracking using recursive
  {G}aussian processes,'' in \emph{2018 21st International Conference on
  Information Fusion (FUSION)}.\hskip 1em plus 0.5em minus 0.4em\relax IEEE,
  2018, pp. 1--8.

\bibitem{baum2009random}
M.~Baum and U.~D. Hanebeck, ``Random hypersurface models for extended object
  tracking,'' in \emph{2009 IEEE International Symposium on Signal Processing
  and Information Technology (ISSPIT)}.\hskip 1em plus 0.5em minus 0.4em\relax
  IEEE, 2009, pp. 178--183.

\bibitem{Tuncer2022_MultiEllipse}
B.~Tuncer, U.~Orguner, and E.~{\"O}zkan, ``Multi-ellipsoidal extended target
  tracking with variational {B}ayes inference,'' \emph{IEEE Transactions on
  Signal Processing}, vol.~70, pp. 3921--3934, 2022.

\bibitem{granstrom2011tracking}
K.~Granstr{\"o}m, C.~Lundquist, and U.~Orguner, ``Tracking rectangular and
  elliptical extended targets using laser measurements,'' in \emph{14th
  International Conference on Information Fusion}.\hskip 1em plus 0.5em minus
  0.4em\relax IEEE, 2011, pp. 1--8.

\bibitem{Michaelis2022}
M.~Michaelis, P.~Berthold, T.~Luettel, and H.-J. Wuensche, ``Extended target
  tracking with a particle filter using state dependent target measurement
  models,'' in \emph{2022 25th International Conference on Information Fusion
  (FUSION)}.\hskip 1em plus 0.5em minus 0.4em\relax IEEE, 2022, pp. 1--8.

\bibitem{Koch2022_VMM}
P.~Hoher, S.~Wirtensohn, T.~Baur, J.~Reuter, F.~Govaers, and W.~Koch,
  ``Extended target tracking with a lidar sensor using random matrices and a
  virtual measurement model,'' \emph{IEEE Transactions on Signal Processing},
  vol.~70, pp. 228--239, 2021.

\bibitem{KochHoher2023_VMM_GP}
P.~Hoher, J.~Reuter, D.~Dold, D.~Griesser, F.~Govaers, and W.~Koch, ``Extended
  target tracking with a lidar sensor using random matrices and a {G}aussian
  processes regression model,'' in \emph{2023 26th International Conference on
  Information Fusion (FUSION)}, 2023, pp. 1--8.

\bibitem{Gridgeman1970_LameOvals}
\BIBentryALTinterwordspacing
N.~T. Gridgeman, ``{Lamé Ovals},'' \emph{The Mathematical Gazette}, vol.~54,
  no. 387, pp. 31--37, 1970. [Online]. Available:
  \url{http://www.jstor.org/stable/3613154}
\BIBentrySTDinterwordspacing

\bibitem{doucet2000sequential}
A.~Doucet, S.~Godsill, and C.~Andrieu, ``On sequential {M}onte {C}arlo sampling
  methods for {B}ayesian filtering,'' \emph{Statistics and computing}, vol.~10,
  pp. 197--208, 2000.

\bibitem{gong2004parametric}
L.~Gong, S.~D. Pathak, D.~R. Haynor, P.~S. Cho, and Y.~Kim, ``Parametric shape
  modeling using deformable superellipses for prostate segmentation,''
  \emph{IEEE transactions on medical imaging}, vol.~23, no.~3, pp. 340--349,
  2004.

\bibitem{Kohntopp2019_SuperellipseMine}
D.~K{\"o}hntopp, B.~Lehmann, D.~Kraus, and A.~Birk, ``Classification and
  localization of naval mines with superellipse active contours,'' \emph{IEEE
  Journal of Oceanic Engineering}, vol.~44, no.~3, pp. 767--782, 2018.

\bibitem{Duchemin2017_SuperellipseContact}
M.~Duchemin, C.~Tugui, and V.~Collee, ``Optimization of contact profiles using
  super-ellipse,'' \emph{SAE International Journal of Materials and
  Manufacturing}, vol.~10, no.~2, pp. 234--244, 2017.

\bibitem{Allen2009_SuperellipseSource}
J.~Allen, N.~Kundtz, D.~A. Roberts, S.~A. Cummer, and D.~R. Smith,
  ``Electromagnetic source transformations using superellipse equations,''
  \emph{Applied Physics Letters}, vol.~94, no.~19, 2009.

\bibitem{kalman1960new}
\BIBentryALTinterwordspacing
R.~E. Kalman, ``{A New Approach to Linear Filtering and Prediction Problems},''
  \emph{Journal of Basic Engineering}, vol.~82, no.~1, pp. 35--45, 03 1960.
  [Online]. Available: \url{https://doi.org/10.1115/1.3662552}
\BIBentrySTDinterwordspacing

\bibitem{Schon2005}
T.~Schon, F.~Gustafsson, and P.-J. Nordlund, ``Marginalized particle filters
  for mixed linear/nonlinear state-space models,'' \emph{IEEE Transactions on
  signal processing}, vol.~53, no.~7, pp. 2279--2289, 2005.

\bibitem{alexe2012measuring}
B.~Alexe, T.~Deselaers, and V.~Ferrari, ``Measuring the objectness of image
  windows,'' \emph{IEEE transactions on pattern analysis and machine
  intelligence}, vol.~34, no.~11, pp. 2189--2202, 2012.

\end{thebibliography}

\end{document}